\documentclass[aps,prl,reprint,superscriptaddress,floatfix,preprintnumbers]{revtex4-2}

\pdfoutput=1

\usepackage[T1]{fontenc}
\usepackage[utf8]{inputenc}
\usepackage{amsmath,amssymb,amsfonts,mathtools,bm}
\usepackage[colorlinks=true, citecolor=blue, linkcolor=blue]{hyperref}
\usepackage[final]{graphicx}
\usepackage{xcolor}
\usepackage{booktabs}
\usepackage{relsize}
\usepackage{lineno}
\usepackage{comment}
\usepackage{braket}
\usepackage{dsfont}
\usepackage{float}
\usepackage{hyperref}
\usepackage{needspace}
\usepackage{enumitem}
\usepackage{braket}

\usepackage{lineno}

\usepackage{csquotes}

\usepackage{algorithm}
\usepackage{algpseudocode}

\def\Babar{{\mbox{\slshape B\kern-0.1em{\smaller A}\kern-0.1em B\kern-0.1em{\smaller A\kern-0.2em R}}}}

\usepackage{cancel}

\usepackage{soul}

\usepackage{siunitx} 
\usepackage[caption=false]{subfig}

\setstcolor{red}

\begin{document}

\preprint{TUM-EFT 220/26}

\title{System-size
dependence of bottomonium suppression from Pb-Pb to light-ion collisions}

\author{Nora Brambilla}
\email{nora.brambilla@tum.de}
\affiliation{Technical University of Munich, TUM School of Natural Sciences, Physics Department, James-Franck-Strasse 1, 85748 Garching, Germany}
\affiliation{Institute for Advanced Study, Technische Universität München, Lichtenbergstrasse 2 a, 85748 Garching, Germany}
\affiliation{Munich Data Science Institute, Technische Universität München, Walther-von-Dyck-Strasse 10, 85748 Garching, Germany}

\author{Tom Magorsch}
\email{tom.magorsch@tum.de}
\affiliation{Technical University of Munich, TUM School of Natural Sciences, Physics Department, James-Franck-Strasse 1, 85748 Garching, Germany}

\author{Antonio Vairo}
\email{antonio.vairo@tum.de}
\affiliation{Technical University of Munich, TUM School of Natural Sciences, Physics Department, James-Franck-Strasse 1, 85748 Garching, Germany}

\begin{abstract}
We test the system-size dependence of bottomonium suppression using the same open-quantum-system transport framework for Pb-Pb, O-O, and Ne-Ne collisions. The microscopic transport coefficients are retained from previous Pb-Pb analyses. Using QTraj, we solve the next-to-leading-order pNRQCD Lindblad equation in anisotropic hydrodynamic backgrounds and calculate integrated $\Upsilon$ double ratios. The QGP-induced evolution reproduces the observed sequential hierarchy and overall magnitude of the O-O suppression measured by CMS and LHCb. We quantify the sensitivity to both transport
coefficients and to the termination temperature of the in-medium
evolution. The Ne-Ne measurement favors somewhat stronger suppression than predicted, although the present experimental uncertainty is large. The extension from Pb-Pb to light ions supports, at least for observables less sensitive to cold nuclear matter effects,
a common microscopic origin of bottomonium suppression in deconfined matter, characterized by the same temperature-dependent transport coefficients despite the large change in collision-system size.
\end{abstract}

\maketitle

\textbf{\textit{Introduction.---}} 
The suppression of heavy quarkonium is a key signature for the existence of a deconfined quark-gluon plasma (QGP) in heavy-ion collisions. Matsui and Satz~\cite{Matsui:1986dk} proposed that the heavy quarkonium potential would be screened by the medium leading to distinct temperatures, above which the bound states would dissolve. This screening would then cause less quarkonia measured in heavy-ion collisions compared to proton-proton collisions. This static picture has been superseded by a dynamical description of heavy quarkonium transport and dissociation in the QGP~\cite{Andronic:2024oxz,Rothkopf:2019ipj}.

In recent years, the description of this process using non-relativistic effective field theories and the open-quantum-system formulation has gained traction~\cite{Yao:2021lus,Akamatsu:2020ypb}. In this framework, the real-time evolution of the quarkonium is described by a quantum master equation. Bottomonium is particularly suitable as its heavy quark mass and small radius provide a strong separation of scales, positioning it as an ideal probe of the QGP. Effective field theories, especially potential non-relativistic QCD (pNRQCD), make it possible to exploit this hierarchy of scales. Using pNRQCD together with the open-quantum-system framework, a quantum master equation for bottomonium has been derived~\cite{Brambilla:2016wgg,Brambilla:2017zei,Brambilla:2022ynh}. This master equation is thus obtained directly from first principles and describes the real-time evolution of the bottomonium in the QGP while retaining its quantum mechanical nature. Under some conditions, the effects of the medium enter through two transport coefficients $\kappa$ and $\gamma$, which are non-perturbative objects with a field-theoretic definition in terms of chromoelectric correlators. The open-source code QTraj was developed and enables the efficient simulation of the master equation~\cite{Omar:2021kra}. By coupling QTraj to a hydrodynamic background, this setup can describe a variety of observables in lead-lead collisions~\cite{Brambilla:2023hkw,Strickland:2023nfm,Brambilla:2024tqg}.

Light-ion collisions provide a controlled testbed to study the system-size dependence of collective QCD phenomena and the relationship between QGP signatures and the size and lifetime of the medium. The 2025 LHC light-ion run delivered O-O and Ne-Ne collisions at $\sqrt{s_{NN}}=5.36\,\mathrm{TeV}$. Bottomonium, once again, suggests itself as a strong probe of the formation of a QGP in these collisions and allows to access the temperature evolution through the sequential suppression of the $\Upsilon(1S,2S,3S)$ states.
Recent measurements of bottomonium suppression in O-O and Ne-Ne collisions by CMS~\cite{CMS:2026qbb} and LHCb~\cite{LHCb:2026jgc} found a pattern of sequential suppression of the excited states with respect to proton-proton collisions. These measurements motivate studying whether in the QTraj framework a hot medium with the temperature and lifetime expected in light-ion collisions can generate the observed state-dependent suppression.

While this work was being completed, Ref.~\cite{Thapa:2026kok} presented a study of bottomonium production in O-O collisions using NLO QTraj and semiclassical transport, including cold-nuclear-matter effects. In its QTraj construction, these effects are modeled by a state-independent factor that cancels in fixed-bin double ratios.

\begin{figure*}[!t]  
    \centering
    \includegraphics[width=\linewidth]{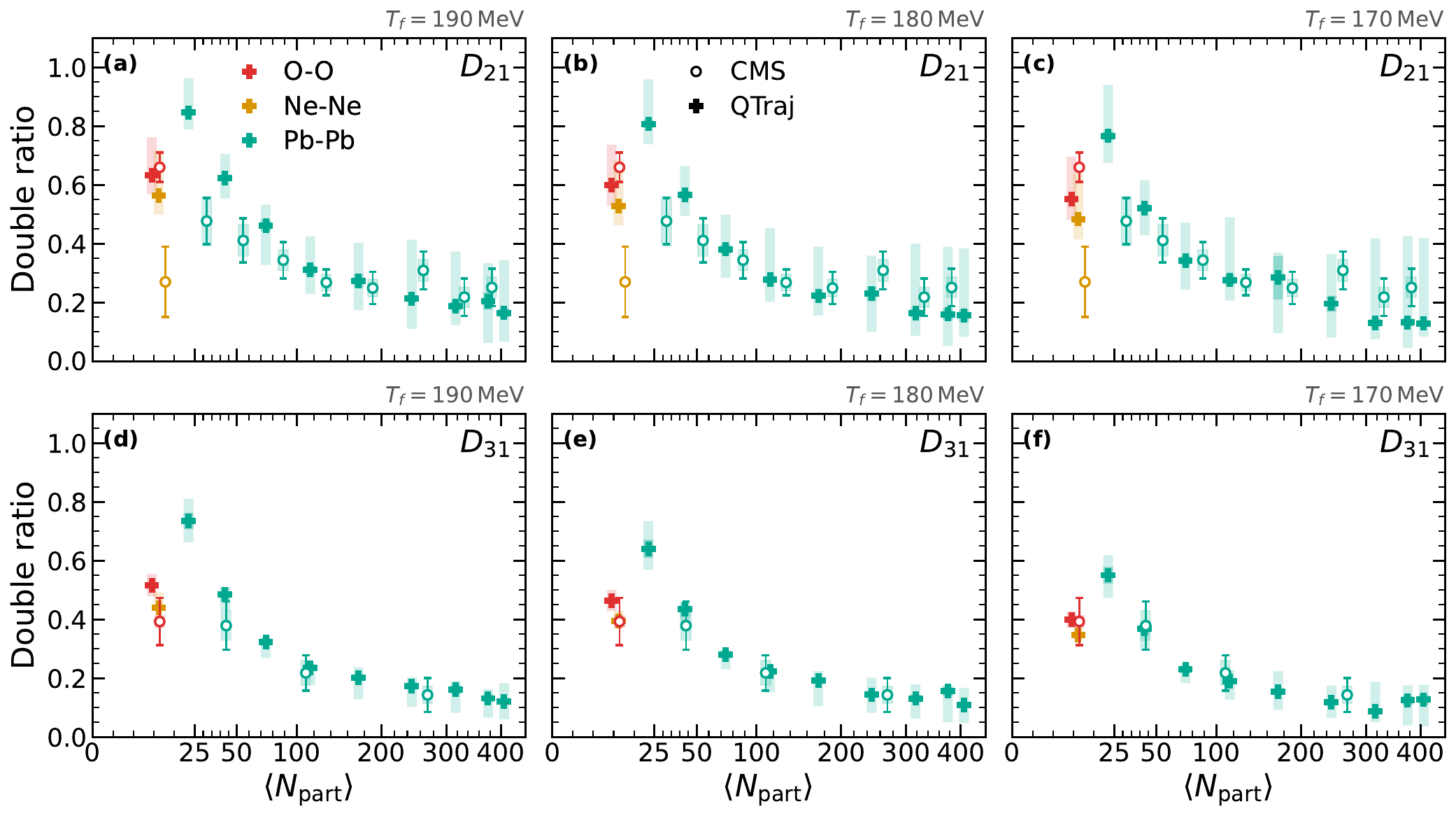}
    \caption{QTraj results for the integrated double ratios $D_{21}$ (top) and $D_{31}$ (bottom) for O-O, Ne-Ne and Pb-Pb collisions compared to CMS measurements~\cite{CMS:2026qbb,CMS:2023lfu,CMS:2018zza} at $\sqrt{s_{NN}}=5.36\,$TeV and $5.02\,$TeV, respectively, as well as $0\leq p_T < 30\,\mathrm{GeV}$ and $|y|<2.4$. Experimental error bars and boxes denote statistical and systematic uncertainties, respectively. The central QTraj prediction corresponds to $\hat\kappa=4$ and $\hat\gamma=0$. The light shaded bands show the independent variation of $\hat\kappa$ and $\hat\gamma$ by $\pm 1$, while the dark shaded bands around the central QTraj prediction show the statistical Monte Carlo uncertainty. The columns show $T_f=190$, $180$, and $170\,$MeV from left to right. The rows show $D_{21}$ and $D_{31}$, respectively. For the minimum-bias O-O and Ne-Ne points, $\braket{N_{\rm part}}$ is defined as the $0$-$100\%$ average over centrality classes. The horizontal axis uses a square-root scale.}
    \label{fig:system_size_comp}
\end{figure*}

In this work we test whether the pNRQCD open-quantum-system description of bottomonium suppression established in Pb-Pb collisions extends to O-O and Ne-Ne collisions without retuning the microscopic transport coefficients to light-ion
bottomonium data. Using the setup established in Pb-Pb studies, we compute the integrated double ratios, compare them with CMS and LHCb measurements, and quantify
the sensitivity to both transport coefficients and to the termination
temperature of the in-medium evolution. 
As shown in Fig.~\ref{fig:system_size_comp}, using the same transport coefficients provides a reasonable description of bottomonium suppression across Pb-Pb, O-O and Ne-Ne collisions within the present uncertainties.

\textbf{\textit{Quarkonium in a medium as an open quantum system.---}}
To describe the evolution of bottomonium in the QGP, the hierarchy of scales can be exploited using non-relativistic effective field theories. The small radius of bottomonium makes the application of pNRQCD particularly favorable. Using pNRQCD, Refs.~\cite{Brambilla:2016wgg,Brambilla:2017zei} derived a quantum master equation for the non-equilibrium evolution of bottomonium in a color medium. By assuming that the binding energy $E$ is smaller than the thermal scale $\pi T$, this equation can be brought into Lindblad form~\cite{Lindblad:1975ef,Gorini:1975nb}, by expanding in $E/(\pi T)$. Refs.~\cite{Brambilla:2016wgg,Brambilla:2017zei} derived the Lindblad equation at leading order in the $E/(\pi T)$ expansion, while Ref.~\cite{Brambilla:2022ynh} derived the next-to-leading-order corrections. In this work, we consider the next-to-leading-order (NLO) master equation. At NLO, the real-time evolution of the quarkonium density matrix $\rho(t)$ can thus be written as
\begin{equation}
    \frac{\mathrm{d}\rho}{\mathrm{d}t} = -i[H,\rho] + \sum^5_{n=0}\left(L_n\rho L^\dagger_n - \frac{1}{2}\left\{L^\dagger_nL_n,\rho\right\} \right).
    \label{eq:lind}
\end{equation}
Here $\rho=\mathrm{diag}(\rho_s,\rho_o)$ is block diagonal in color, with $\rho_s$ and $\rho_o$ denoting the color-singlet and color-octet components, respectively. Both $\rho_s$ and $\rho_o$ are written in the spherical representation and are block diagonal in the angular momentum $\ell$, such that $\rho_c=\bigoplus_\ell\rho_c^{(\ell)}$, where $\rho_c^{(\ell)}$ is the density matrix in the angular-momentum sector $\ell$. Furthermore, $H$ denotes the Hamiltonian, while the collapse operators $L_n$ parametrize the QGP-induced dissipation. We provide the explicit form of the operators in the Supplemental Material. 

Under some simplifying assumptions (neglecting effects due to different Wilson line arrangements and using the instantaneous approximation for the first moment correlator), the Hamiltonian and collapse operators depend on two transport coefficients, the heavy quarkonium diffusion coefficient $\kappa$ and its dispersive counterpart $\gamma$. We write both in terms of dimensionless parameters $\kappa=\hat\kappa T^3$ and $\gamma=\hat\gamma T^3$. These coefficients are non-perturbative objects defined in terms of chromoelectric correlators that parametrize the properties of the QGP. Within the local-equilibrium approximation and at fixed thermodynamic conditions, these coefficients characterize the medium rather than the nuclei that produced it. We therefore use the same functions $\kappa$ and $\gamma$ in all collision systems. Approximating the dimensionless coefficients
$\hat\kappa$ and $\hat\gamma$ as constants is an additional modeling assumption, which we also keep unchanged. While the diffusion coefficient for the transport of a heavy quark in the QGP has been calculated on the lattice~\cite{Francis:2015daa,Brambilla:2020siz,Altenkort:2023oms,HotQCD:2025fbd}, first lattice determinations of the corresponding heavy-quarkonium transport coefficients are now underway~\cite{Brambilla:2025cqy}. On the other hand, Ref.~\cite{Brambilla:2023hkw} found that the constant dimensionless transport coefficients $\hat\kappa=4$ and $\hat\gamma=0$ describe the measured bottomonium suppression in Pb-Pb collisions well.

The solution of the Lindblad equation~\eqref{eq:lind} has been implemented in the open-source code QTraj~\cite{Omar:2021kra}, which enables an efficient unraveling of the full master equation by the Monte Carlo wavefunction method~\cite{PhysRevLett.68.580}. Coupling the simulation of the bottomonium evolution through the Lindblad equation to hydrodynamics, which supplies the temperature evolution of the medium, enabled phenomenological predictions for different observables at the LHC and RHIC~\cite{Brambilla:2023hkw,Brambilla:2024tqg,Strickland:2023nfm,Strickland:2024oat}. We here use this setup to make predictions for bottomonium suppression in O-O and Ne-Ne collisions at the LHC.

\textbf{\textit{Hydrodynamics.---}}
In this work we test the framework previously established for Pb-Pb collisions~\cite{Brambilla:2020qwo,Brambilla:2021wkt,Brambilla:2022ynh,Brambilla:2023hkw}.
To this end, we employ the same $(3+1)$-dimensional anisotropic-hydrodynamics framework based on aHydroQP~\cite{Alqahtani:2015qja,Alqahtani:2016rth,Alqahtani:2017mhy,Alqahtani:2020paa}, closely following the setup of the previous QTraj-based Pb-Pb studies.
We retain the smooth optical-Glauber initial conditions used in these calculations. Event-by-event fluctuations, neglected here, may become more important in a light system such as O-O, especially for peripheral collisions, although previous Pb-Pb studies found only a small effect on inclusive bottomonium suppression~\cite{Alalawi:2022gul}.

The hydrodynamic evolution is initialized at $\tau_0=0.25\,\mathrm{fm}$ for O-O collisions at $\sqrt{s_{NN}}=5.36\,\mathrm{TeV}$.
We keep the shear-viscosity-to-entropy-density ratio fixed at $\eta/s=0.159$, as in previous Pb-Pb tunes~\cite{Alqahtani:2017tnq,Alqahtani:2020paa}.
In contrast to previous Pb-Pb calculations, we initialize the system with a finite momentum-space anisotropy, $(\alpha_x,\alpha_y,\alpha_z)=(1,1,0.4)$, estimated from free streaming between a microscopic formation time $\tau_\mathrm{micro}\simeq0.1\,\mathrm{fm}$ and $\tau_0$~\cite{Martinez_2008,Strickland_2014}. The parameters of the smooth optical-Glauber initial conditions are then tuned to reproduce the centrality-integrated charged-particle pseudorapidity distribution measured by CMS~\cite{CMS:2026sai}, giving an initial central effective temperature of $T_0=466\,\mathrm{MeV}$. Here $T_0$ is the effective temperature obtained by Landau matching the energy density of the anisotropic distribution to that of an equilibrium one. Further details of the hydrodynamic setup are given in the Supplemental Material.

\begin{figure}[t]
    \centering
    \includegraphics[width=0.9\linewidth]{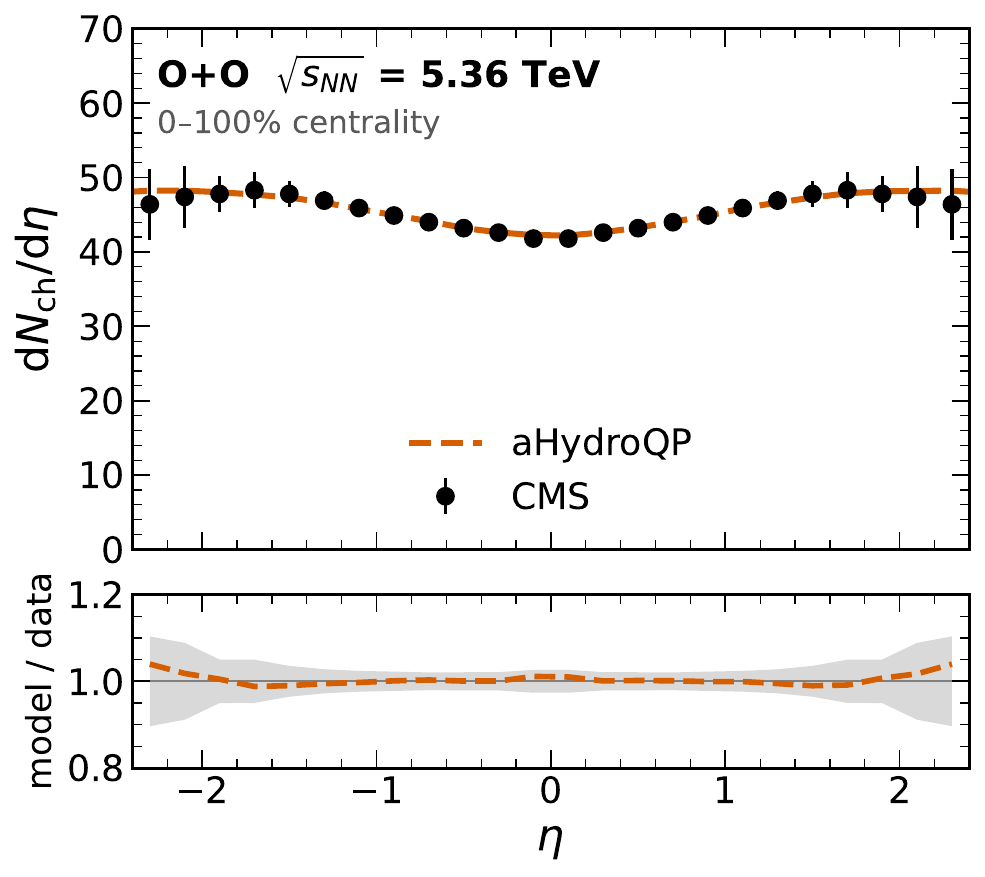}
    \caption{The charged-particle multiplicity in terms of pseudorapidity $\eta$ in $0$-$100\%$ centrality for O-O collisions, compared with data from the CMS experiment~\cite{CMS:2026sai}.}
    \label{fig:hydro}
\end{figure}

As in previous studies, we use a version of the THERMINATOR~2 code~\cite{Chojnacki:2011hb} to simulate the production and decay of primordial hadrons.
The resulting charged-particle pseudorapidity distribution is shown in Fig.~\ref{fig:hydro} and reproduces the CMS data well. We discuss the resulting centrality dependence of the charged-particle multiplicity and of the mean transverse momentum together with the corresponding limitations of the hydrodynamic description in the Supplemental Material.

To study Ne-Ne collisions, we keep the setup identical to the oxygen case. Instead of introducing a separate parametrization of the neon nuclear density, we only change the mass number to $A=20$, while retaining the same Woods-Saxon parametrization and Glauber initial conditions. Within this construction the larger mass number leads to a larger initialized energy density. Using the rescaled energy density, we then fix the initial temperature of the Glauber initial conditions using the equation of state. We obtain the temperature $T_0$ corresponding to the rescaled neon energy density as $T^\mathrm{Ne}_0=481\,\mathrm{MeV}$. We use this hydrodynamic background for the Ne-Ne collisions.

\textbf{\textit{Bottomonium suppression.---}}
We use the hydrodynamic background to sample about $\num{3.5e5}$ physical bottomonium trajectories through the plasma in the same manner as previous lead-lead studies~\cite{Brambilla:2020qwo,Brambilla:2021wkt,Brambilla:2022ynh,Brambilla:2023hkw}. We then record the temperature evolution of each physical trajectory as it traverses the plasma. We initialize the evolution of the Lindblad equation ~\eqref{eq:lind} at $\tau_\text{med}=0.6\,$fm. Using QTraj, we then sample about $30$ quantum trajectories for each physical trajectory, using its respective temperature evolution. Following past studies, we terminate the QTraj evolution at a final temperature of $T_f$~\cite{Brambilla:2023hkw}. We take
$T_f=190\,$MeV as the reference termination temperature, as in the preceding Pb-Pb analysis~\cite{Brambilla:2022ynh,Brambilla:2023hkw}, and use the same value in all collision systems. To assess sensitivity to the late-time evolution, we also perform calculations with
$T_f=180\,$MeV and $170\,$MeV. These variations probe the termination prescription and the extrapolation of the binding-energy expansion toward lower temperatures. For the setup of the QTraj code we adopt the choices of previous studies~\cite{Brambilla:2022ynh,Brambilla:2023hkw}, which we give in the Supplemental Material.
After terminating the QTraj evolution, we record the survival probabilities of the $\Upsilon(1S)$, $\Upsilon(2S)$, $\Upsilon(3S)$, $\chi_b(1P)$ and $\chi_b(2P)$ states as the projection of the density matrix on the respective vacuum eigenstate normalized by the initial overlaps. We then apply an excited-state-feed-down using branching-fraction data from the Particle Data Group~\cite{ParticleDataGroup:2026mpi}, by introducing the vector of experimentally observed $pp$ production cross sections $\boldsymbol{\sigma}_{\mathrm{exp}}^{pp}$ and the corresponding vector of direct production cross sections $\boldsymbol{\sigma}_{\mathrm{direct}}^{pp}$ for the considered states. These two cross sections are related by $\boldsymbol{\sigma}_{\mathrm{exp}}^{pp}= F\boldsymbol{\sigma}_{\mathrm{direct}}^{pp}$ and $\boldsymbol{\sigma}_{\mathrm{direct}}^{pp}=F^{-1}\boldsymbol{\sigma}_{\mathrm{exp}}^{pp}$, where $F$ is the feed-down matrix with the matrix elements given by the PDG branching fractions. We use the same experimental input cross sections and feed-down matrix as in Ref.~\cite{Brambilla:2020qwo}. The feed-down-corrected nuclear modification factor at a fixed centrality $c$ is then calculated as
\begin{equation}
R_{AA}^{i}(c)
    =
    \frac{
        \left[F S(c)\boldsymbol{\sigma}_{\mathrm{direct}}^{pp}\right]_i
    }{
        \left[\boldsymbol{\sigma}_{\mathrm{exp}}^{pp}\right]_i
    },
    \label{eq:raa}
\end{equation} 
where $S(c)$ is a diagonal matrix containing the trajectory-averaged QTraj survival probabilities for the given centrality. 
\begin{figure*}[!t]
    \includegraphics[width=0.9\linewidth]{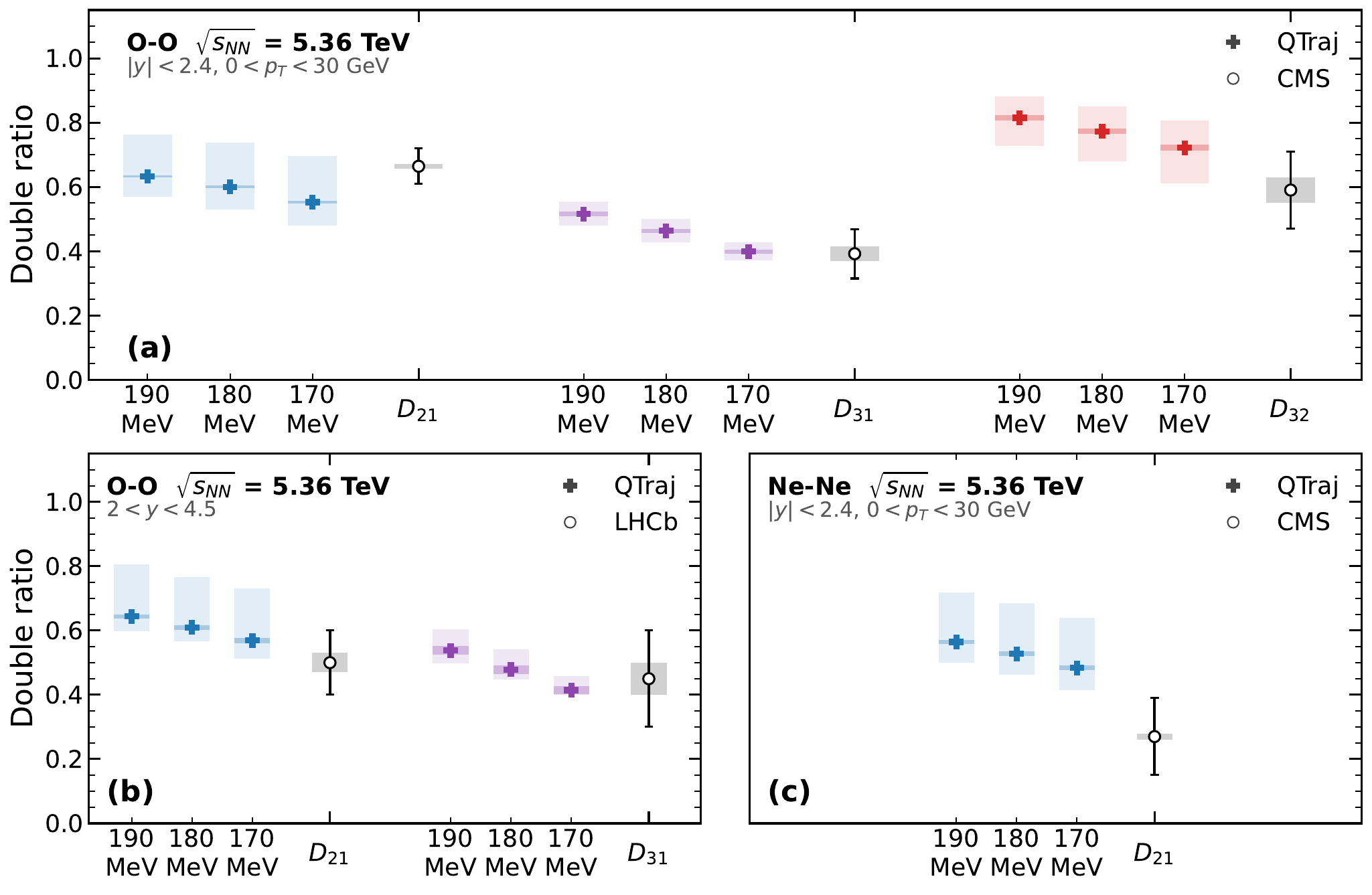}
    \caption{QTraj results for the integrated double ratios $D_{ij}$. Panel (a) compares to the O-O measurement by CMS~\cite{CMS:2026qbb}, panel (b) compares to the O-O measurement by LHCb~\cite{LHCb:2026jgc} and panel (c) to the Ne-Ne measurement by CMS~\cite{CMS:2026qbb}. The central QTraj prediction corresponds to $\hat\kappa=4$ and $\hat\gamma=0$. The light shaded bands show the independent variation of $\hat\kappa$ and $\hat\gamma$ by $\pm 1$, while the dark shaded bands around the central QTraj predictions show the statistical Monte Carlo uncertainty. We compare the QTraj result for different final temperatures $T_f$ of the simulation.}
    \label{fig:integrated}
\end{figure*}
The centrality-integrated double ratios $D_{ij}$ between two states $\Upsilon(i\mathrm{S})$ and $\Upsilon(j\mathrm{S})$ are then calculated as 
\begin{equation}
    D_{ij}=\frac{\sum_k \Delta c_k \braket{N_\text{coll}}_k R^i_{AA}(c_k)}{\sum_k \Delta c_k \braket{N_\text{coll}}_k R^j_{AA}(c_k)}.
    \label{eq:double_ratio}
\end{equation}
Here $\Delta c_k$ denotes the relative width of the centrality class $c_k$ and $\braket{N_\text{coll}}_k$ the respective count of binary collisions obtained from the smooth optical-Glauber model. The weight $\Delta c_k \braket{N_\text{coll}}_k$ accounts for the  probability of the centrality class and the assumed binary-collision scaling of hard bottomonium production. Our predictions include hot-medium evolution but not cold-nuclear-matter effects. Double ratios reduce sensitivity to approximately state-independent modifications of the initial production, such as those modeled in Ref.~\cite{Thapa:2026kok}. Residual effects from centrality and kinematic reweighting are not quantified here.

For the transport coefficients we use the central values $\hat\kappa=4$ and $\hat\gamma=0$, which were previously found to describe the experimental data in Pb-Pb collisions well~\cite{Brambilla:2023hkw}. To account for the uncertainty of this choice, we take the envelope of the nine predictions from independently varying $\hat\kappa=4\pm1$ and $\hat\gamma=0\pm 1$.

Figure~\ref{fig:system_size_comp} compares light-ion and Pb-Pb predictions with respective measurements by CMS~\cite{CMS:2023lfu,CMS:2018zza,CMS:2026qbb}. For the minimum-bias O-O and Ne-Ne points we use $\braket{N_\text{part}}=\left(\sum_k\Delta c_k \braket{N_\text{part}}_k\right)/\left(\sum_k \Delta c_k\right)$ consistent with the minimum-bias averaging used experimentally. The Pb-Pb results are computed with the hydrodynamic setup from previous studies~\cite{Brambilla:2023hkw,Strickland:2023nfm}. In all collision systems we use the same transport coefficients, while the corresponding hydrodynamic backgrounds account for the different medium evolutions. 
The measured suppression is reasonably described across collision systems for all three choices of $T_f$. Lowering $T_f$ leads to systematically stronger suppression, with the sensitivity becoming most pronounced toward the light-ion and peripheral Pb-Pb systems. We emphasize, however, that the O-O and Ne-Ne results are centrality integrated, whereas the Pb-Pb points correspond to individual centrality classes. We compare the transverse momentum dependence of the double ratios to measurements by CMS in the Supplemental Material.

In Fig.~\ref{fig:integrated} we show predictions for the integrated double ratios compared to CMS~\cite{CMS:2026qbb} and LHCb~\cite{LHCb:2026jgc} measurements. We notice that the CMS and LHCb measurements were performed in different rapidity windows. In panel (a), we find that the $D_{21}$ and $D_{31}$ double ratios for O-O collisions are compatible with the CMS measurement, while the predicted $D_{32}$ is slightly larger than the measured central value. Comparing different choices of the final temperature $T_f$, decreasing $T_f$ leads to stronger suppression, as the bottomonium spends a longer time in the QGP. 
For $D_{31}$ and $D_{32}$, lower $T_f$ values shift the predictions toward the CMS central values. In panel (b) we find that our results are likewise compatible with the LHCb measurements in O-O collisions, with lower $T_f$ reducing the difference in $D_{21}$. In particular, the QTraj evolution reproduces the observed hierarchy between the bottomonium states using only hot-medium effects. This shows that the temperature and lifetime reached in O-O collisions are sufficient to generate sizable state-dependent bottomonium suppression and supports the interpretation that a QGP is formed in these collisions.

In panel (c) of  Fig.~\ref{fig:integrated} we compare the integrated double ratio $D_{21}$ for Ne-Ne collisions with the respective CMS measurement. The predicted $D_{21}$ is approximately $11$-$13\%$ smaller in Ne-Ne than in O-O over the range $T_f=170$-$190\,$MeV. Although the individual double ratios depend appreciably on the termination temperature, their ratio changes
comparatively little. The central CMS value indicates a stronger difference between the two systems, although the large uncertainty of the Ne-Ne measurement prevents a firm conclusion. Within our hydrodynamic setup, the increase in system size from oxygen to neon is therefore not sufficient to produce a large change in $D_{21}$. If a substantially stronger separation between O-O and Ne-Ne is confirmed by more precise data, it would point towards either a stronger change in the medium evolution than captured by the present smooth hydrodynamic backgrounds or additional nuclear effects not included in the calculation.

\textbf{\textit{Conclusions.---}}
We have tested the NLO pNRQCD open-quantum-system description of
bottomonium suppression across Pb-Pb, O-O and Ne-Ne collisions. The
reference transport coefficients, $\hat\kappa=4$ and $\hat\gamma=0$,
were retained from the previous Pb-Pb analysis without refitting them to
light-ion bottomonium data. The resulting QGP-induced evolution reproduces the sequential hierarchy and overall magnitude of the O-O double ratios measured by CMS and LHCb. The description established in Pb-Pb therefore has predictive reach in a substantially smaller collision system, without introducing new microscopic transport parameters.

We quantified the sensitivity to both transport coefficients and to the
termination temperature of the in-medium evolution. For the reference
coefficients, the predicted $D_{21}$ is approximately $11$-$13\%$ smaller in Ne-Ne than in O-O throughout $T_f=170$-$190\,$MeV. Thus, varying the termination temperature changes the absolute double ratios more strongly than the relative separation between the O-O and Ne-Ne systems. The larger separation suggested by the CMS central values is not reproduced by this variation alone. A further source of uncertainty is the use of smooth optical-Glauber initial conditions, and future studies with event-by-event fluctuating initial conditions will be important for a more quantitative description of these small collision systems.

For the double ratios studied here, which have reduced sensitivity to approximately state-independent cold-nuclear-matter effects, the extension from Pb-Pb to O-O supports a common microscopic origin of bottomonium suppression in deconfined matter. Within the local-equilibrium description, the same temperature-dependent transport coefficients characterize the response of the plasma, while the different medium histories generate the collision-system dependence. Light-ion bottomonium measurements therefore provide a nontrivial test of the microscopic QGP dynamics previously constrained in heavy-ion collisions.

\section*{Acknowledgments}
N.B., T.M. and A.V. thank Soohwan Lee for useful discussion.
N.B., T.M. and A.V. thank Michael Strickland for providing the aHydroQP code and instructions for its use. N.B., T.M. and A.V. acknowledge support by the DFG cluster of excellence ORIGINS funded by the Deutsche Forschungsgemeinschaft (DFG) under Germany's Excellence Strategy - EXC-2094-390783311. The work of N.B.
is supported by the DFG Grant No. BR 4058/5-1 ”Open quantum systems and effective field theories for hard probes of hot and/or dense medium”. N.B. acknowledges the European Research Council advanced grant ERC-2023-ADG-Project EFT-XYZ.

\bibliographystyle{apsrev4-1}
\renewcommand*{\bibfont}{\footnotesize}
\bibliography{lit.bib}

@article{Brambilla:2023hkw,
    author = "Brambilla, Nora and Escobedo, Miguel {\'A}ngel and Islam, Ajaharul and Strickland, Michael and Tiwari, Anurag and Vairo, Antonio and Vander Griend, Peter",
    title = "{Regeneration of bottomonia in an open quantum systems approach}",
    eprint = "2302.11826",
    archivePrefix = "arXiv",
    primaryClass = "hep-ph",
    reportNumber = "TUM-EFT 178/23, FERMILAB-PUB-23-060-V, TUM-EFT 178/23; FERMILAB-PUB-23-060-V",
    doi = "10.1103/PhysRevD.108.L011502",
    journal = "Phys. Rev. D",
    volume = "108",
    number = "1",
    pages = "L011502",
    year = "2023"
}

@article{Brambilla:2022ynh,
    author = "Brambilla, Nora and Escobedo, Miguel {\'A}ngel and Islam, Ajaharul and Strickland, Michael and Tiwari, Anurag and Vairo, Antonio and Vander Griend, Peter",
    title = "{Heavy quarkonium dynamics at next-to-leading order in the binding energy over temperature}",
    eprint = "2205.10289",
    archivePrefix = "arXiv",
    primaryClass = "hep-ph",
    reportNumber = "TUM-EFT 169/22",
    doi = "10.1007/JHEP08(2022)303",
    journal = "JHEP",
    volume = "08",
    pages = "303",
    year = "2022",
    note = "[Erratum: JHEP 11, 079 (2025)]"
}

@article{Brambilla:2021wkt,
    author = "Brambilla, Nora and Escobedo, Miguel {\'A}ngel and Strickland, Michael and Vairo, Antonio and Vander Griend, Peter and Weber, Johannes Heinrich",
    title = "{Bottomonium production in heavy-ion collisions using quantum trajectories: Differential observables and momentum anisotropy}",
    eprint = "2107.06222",
    archivePrefix = "arXiv",
    primaryClass = "hep-ph",
    reportNumber = "TUM-EFT 147/21, HU-EP-21/18-RTG",
    doi = "10.1103/PhysRevD.104.094049",
    journal = "Phys. Rev. D",
    volume = "104",
    number = "9",
    pages = "094049",
    year = "2021"
}

@article{Brambilla:2020qwo,
    author = "Brambilla, Nora and Escobedo, Miguel {\'A}ngel and Strickland, Michael and Vairo, Antonio and Vander Griend, Peter and Weber, Johannes Heinrich",
    title = "{Bottomonium suppression in an open quantum system using the quantum trajectories method}",
    eprint = "2012.01240",
    archivePrefix = "arXiv",
    primaryClass = "hep-ph",
    reportNumber = "TUM-EFT 140/20; HU-EP-20/36-RTG",
    doi = "10.1007/JHEP05(2021)136",
    journal = "JHEP",
    volume = "05",
    pages = "136",
    year = "2021"
}

@article{Alqahtani:2020paa,
    author = "Alqahtani, Mubarak and Strickland, Michael",
    title = "{Bulk observables at 5.02~TeV using quasiparticle anisotropic hydrodynamics}",
    eprint = "2008.07657",
    archivePrefix = "arXiv",
    primaryClass = "nucl-th",
    doi = "10.1140/epjc/s10052-021-09832-z",
    journal = "Eur. Phys. J. C",
    volume = "81",
    number = "11",
    pages = "1022",
    year = "2021"
}

@article{Alqahtani:2015qja,
    author = "Alqahtani, Mubarak and Nopoush, Mohammad and Strickland, Michael",
    title = "{Quasiparticle equation of state for anisotropic hydrodynamics}",
    eprint = "1509.02913",
    archivePrefix = "arXiv",
    primaryClass = "hep-ph",
    doi = "10.1103/PhysRevC.92.054910",
    journal = "Phys. Rev. C",
    volume = "92",
    number = "5",
    pages = "054910",
    year = "2015"
}

@article{Alqahtani:2016rth,
    author = "Alqahtani, Mubarak and Nopoush, Mohammad and Strickland, Michael",
    title = "{Quasiparticle anisotropic hydrodynamics for central collisions}",
    eprint = "1605.02101",
    archivePrefix = "arXiv",
    primaryClass = "nucl-th",
    doi = "10.1103/PhysRevC.95.034906",
    journal = "Phys. Rev. C",
    volume = "95",
    number = "3",
    pages = "034906",
    year = "2017"
}

@article{Alqahtani:2017mhy,
    author = "Alqahtani, Mubarak and Nopoush, Mohammad and Strickland, Michael",
    title = "{Relativistic anisotropic hydrodynamics}",
    eprint = "1712.03282",
    archivePrefix = "arXiv",
    primaryClass = "nucl-th",
    doi = "10.1016/j.ppnp.2018.05.004",
    journal = "Prog. Part. Nucl. Phys.",
    volume = "101",
    pages = "204--248",
    year = "2018"
}

@article{Chojnacki:2011hb,
    author = "Chojnacki, Mikolaj and Kisiel, Adam and Florkowski, Wojciech and Broniowski, Wojciech",
    title = "{THERMINATOR 2: THERMal heavy IoN generATOR 2}",
    eprint = "1102.0273",
    archivePrefix = "arXiv",
    primaryClass = "nucl-th",
    doi = "10.1016/j.cpc.2011.11.018",
    journal = "Comput. Phys. Commun.",
    volume = "183",
    pages = "746--773",
    year = "2012"
}

@article{Strickland:2023nfm,
    author = "Strickland, Michael and Thapa, Sabin",
    title = "{Bottomonium suppression at RHIC and LHC in an open quantum system approach}",
    eprint = "2305.17841",
    archivePrefix = "arXiv",
    primaryClass = "hep-ph",
    doi = "10.1103/PhysRevD.108.014031",
    journal = "Phys. Rev. D",
    volume = "108",
    number = "1",
    pages = "014031",
    year = "2023"
}

@article{Martinez_2008,
   title={Pre-equilibrium dilepton production from an anisotropic quark-gluon plasma},
   volume={78},
   ISSN={1089-490X},
   url={http://dx.doi.org/10.1103/PhysRevC.78.034917},
   DOI={10.1103/physrevc.78.034917},
   number={3},
   journal={Physical Review C},
   publisher={American Physical Society (APS)},
   author={Martinez, Mauricio and Strickland, Michael},
   year={2008},
   month=sep }

@article{Strickland_2014,
   title={Anisotropic Hydrodynamics: Three Lectures},
   volume={45},
   ISSN={1509-5770},
   url={http://dx.doi.org/10.5506/APhysPolB.45.2355},
   DOI={10.5506/aphyspolb.45.2355},
   number={12},
   journal={Acta Physica Polonica B},
   publisher={Jagiellonian University},
   author={Strickland, M.},
   year={2014},
   pages={2355} }

@article{Loizides:2025ule,
    author = "Loizides, Constantin",
    title = "{Glauber predictions for oxygen and neon collisions at energies available at the CERN Large Hadron Collider}",
    eprint = "2507.05853",
    archivePrefix = "arXiv",
    primaryClass = "nucl-th",
    doi = "10.1103/mkp8-zgxh",
    journal = "Phys. Rev. C",
    volume = "113",
    number = "1",
    pages = "014914",
    year = "2026"}

@article{Alalawi:2022gul,
    author = "Alalawi, Huda and Boyd, Jacob and Shen, Chun and Strickland, Michael",
    title = "{Impact of fluctuating initial conditions on bottomonium suppression in 5.02 TeV heavy-ion collisions}",
    eprint = "2211.06363",
    archivePrefix = "arXiv",
    primaryClass = "hep-ph",
    doi = "10.1103/PhysRevC.107.L031901",
    journal = "Phys. Rev. C",
    volume = "107",
    number = "3",
    pages = "L031901",
    year = "2023"
}

@article{CMS:2026sai,
    author = "Belyaev, Andrey and others",
    collaboration = "CMS",
    title = "{Centrality dependence of charged-hadron pseudorapidity distributions in oxygen-oxygen collisions at $\sqrt{s_\mathrm{NN}}$ = 5.36 TeV}",
    eprint = "2606.02285",
    archivePrefix = "arXiv",
    primaryClass = "nucl-ex",
    reportNumber = "CMS-HIN-25-010, CERN-EP-2026-153",
    month = "6",
    year = "2026"
}

@article{ATLAS:2026zgq,
    author = "Aad, Georges and others",
    collaboration = "ATLAS",
    title = "{Measurements of charged-particle pseudorapidity and transverse momentum distributions in O+O and Ne+Ne collisions at $\sqrt{s_{_\text{NN}}} = 5.36$ TeV with the ATLAS detector}",
    eprint = "2606.20257",
    archivePrefix = "arXiv",
    primaryClass = "nucl-ex",
    reportNumber = "CERN-EP-2026-154",
    month = "6",
    year = "2026"
}

@article{Alqahtani:2017tnq,
    author = "Alqahtani, Mubarak and Nopoush, Mohammad and Ryblewski, Radoslaw and Strickland, Michael",
    title = "{Anisotropic hydrodynamic modeling of 2.76 TeV Pb-Pb collisions}",
    eprint = "1705.10191",
    archivePrefix = "arXiv",
    primaryClass = "nucl-th",
    doi = "10.1103/PhysRevC.96.044910",
    journal = "Phys. Rev. C",
    volume = "96",
    number = "4",
    pages = "044910",
    year = "2017"
}

@article{DEVRIES1987495,
title = {Nuclear charge-density-distribution parameters from elastic electron scattering},
journal = {Atomic Data and Nuclear Data Tables},
volume = {36},
number = {3},
pages = {495-536},
year = {1987},
issn = {0092-640X},
doi = {https://doi.org/10.1016/0092-640X(87)90013-1},
url = {https://www.sciencedirect.com/science/article/pii/0092640X87900131},
author = {H. {De Vries} and C.W. {De Jager} and C. {De Vries}}
}

@article{HotQCD:2014kol,
    author = "Bazavov, A. and others",
    collaboration = "HotQCD",
    title = "{Equation of state in ( 2+1 )-flavor QCD}",
    eprint = "1407.6387",
    archivePrefix = "arXiv",
    primaryClass = "hep-lat",
    reportNumber = "BNL-105928-2014-JA",
    doi = "10.1103/PhysRevD.90.094503",
    journal = "Phys. Rev. D",
    volume = "90",
    pages = "094503",
    year = "2014"
}

@article{Brambilla:2016wgg,
    author = "Brambilla, Nora and Escobedo, Miguel A. and Soto, Joan and Vairo, Antonio",
    title = "{Quarkonium suppression in heavy-ion collisions: an open quantum system approach}",
    eprint = "1612.07248",
    archivePrefix = "arXiv",
    primaryClass = "hep-ph",
    reportNumber = "ICCUB-16-044, TUM-EFT-55-14",
    doi = "10.1103/PhysRevD.96.034021",
    journal = "Phys. Rev. D",
    volume = "96",
    number = "3",
    pages = "034021",
    year = "2017"
}

@article{Brambilla:2017zei,
    author = "Brambilla, Nora and Escobedo, Miguel A. and Soto, Joan and Vairo, Antonio",
    title = "{Heavy quarkonium suppression in a fireball}",
    eprint = "1711.04515",
    archivePrefix = "arXiv",
    primaryClass = "hep-ph",
    reportNumber = "TUM-EFT-89-16",
    doi = "10.1103/PhysRevD.97.074009",
    journal = "Phys. Rev. D",
    volume = "97",
    number = "7",
    pages = "074009",
    year = "2018"
}

@article{Lindblad:1975ef,
    author = "Lindblad, Goran",
    title = "{On the Generators of Quantum Dynamical Semigroups}",
    reportNumber = "TRITA-TFY-75-1",
    doi = "10.1007/BF01608499",
    journal = "Commun. Math. Phys.",
    volume = "48",
    pages = "119",
    year = "1976"
}

@article{Gorini:1975nb,
    author = "Gorini, Vittorio and Kossakowski, Andrzej and Sudarshan, E. C. G.",
    title = "{Completely Positive Dynamical Semigroups of N Level Systems}",
    reportNumber = "CPT-244-TEXAS, ORO-3992-200",
    doi = "10.1063/1.522979",
    journal = "J. Math. Phys.",
    volume = "17",
    pages = "821",
    year = "1976"
}

@article{Brambilla:2020siz,
    author = "Brambilla, Nora and Leino, Viljami and Petreczky, Peter and Vairo, Antonio",
    title = "{Lattice QCD constraints on the heavy quark diffusion coefficient}",
    eprint = "2007.10078",
    archivePrefix = "arXiv",
    primaryClass = "hep-lat",
    reportNumber = "TUM-EFT 131/19",
    doi = "10.1103/PhysRevD.102.074503",
    journal = "Phys. Rev. D",
    volume = "102",
    number = "7",
    pages = "074503",
    year = "2020"
}

@article{HotQCD:2025fbd,
    author = "Bollweg, Dennis and Dasilva Gol{\'a}n, Jorge Luis and Kaczmarek, Olaf and Larsen, Rasmus Norman and Moore, Guy D. and Mukherjee, Swagato and Petreczky, Peter and Shu, Hai-Tao and Stendebach, Simon and Weber, Johannes Heinrich",
    collaboration = "HotQCD",
    title = "{Temperature dependence of heavy quark diffusion from (2+1)-flavor lattice QCD}",
    eprint = "2506.11958",
    archivePrefix = "arXiv",
    primaryClass = "hep-lat",
    doi = "10.1007/JHEP09(2025)180",
    journal = "JHEP",
    volume = "09",
    pages = "180",
    year = "2025"
}

@article{Altenkort:2023oms,
    author = "Altenkort, Luis and Kaczmarek, Olaf and Larsen, Rasmus and Mukherjee, Swagato and Petreczky, Peter and Shu, Hai-Tao and Stendebach, Simon",
    collaboration = "HotQCD",
    title = "{Heavy Quark Diffusion from 2+1 Flavor Lattice QCD with 320~MeV Pion Mass}",
    eprint = "2302.08501",
    archivePrefix = "arXiv",
    primaryClass = "hep-lat",
    doi = "10.1103/PhysRevLett.130.231902",
    journal = "Phys. Rev. Lett.",
    volume = "130",
    number = "23",
    pages = "231902",
    year = "2023"
}

@article{Francis:2015daa,
    author = "Francis, A. and Kaczmarek, O. and Laine, M. and Neuhaus, T. and Ohno, H.",
    title = "{Nonperturbative estimate of the heavy quark momentum diffusion coefficient}",
    eprint = "1508.04543",
    archivePrefix = "arXiv",
    primaryClass = "hep-lat",
    reportNumber = "BI-TP-2015-11, HIP-2015-28-TH",
    doi = "10.1103/PhysRevD.92.116003",
    journal = "Phys. Rev. D",
    volume = "92",
    number = "11",
    pages = "116003",
    year = "2015"
}

@article{Brambilla:2025cqy,
    author = "Brambilla, Nora and Datta, Saumen and Janer, Marc and Leino, Viljami and Mayer-Steudte, Julian and Petreczky, Peter and Vairo, Antonio",
    collaboration = "TUMQCD",
    title = "{Lattice study of correlators of chromoelectric fields for heavy quarkonium dynamics in the quark-gluon plasma}",
    eprint = "2505.16603",
    archivePrefix = "arXiv",
    primaryClass = "hep-lat",
    reportNumber = "TUM-EFT 189/24, MITP-24-077",
    doi = "10.1103/387k-mdl1",
    journal = "Phys. Rev. D",
    volume = "112",
    number = "7",
    pages = "074509",
    year = "2025"
}

@article{Omar:2021kra,
    author = "Omar, Hisham Ba and Escobedo, Miguel {\'A}ngel and Islam, Ajaharul and Strickland, Michael and Thapa, Sabin and Vander Griend, Peter and Weber, Johannes Heinrich",
    title = "{QTRAJ 1.0: A Lindblad equation solver for heavy-quarkonium dynamics}",
    eprint = "2107.06147",
    archivePrefix = "arXiv",
    primaryClass = "physics.comp-ph",
    reportNumber = "TUM-EFT 142/21; HU-EP-21/17-RTG",
    doi = "10.1016/j.cpc.2021.108266",
    journal = "Comput. Phys. Commun.",
    volume = "273",
    pages = "108266",
    year = "2022"
}

@article{PhysRevLett.68.580,
  title = {Wave-function approach to dissipative processes in quantum optics},
  author = {Dalibard, Jean and Castin, Yvan and M\o{}lmer, Klaus},
  journal = {Phys. Rev. Lett.},
  volume = {68},
  issue = {5},
  pages = {580--583},
  numpages = {0},
  year = {1992},
  month = {Feb},
  publisher = {American Physical Society},
  doi = {10.1103/PhysRevLett.68.580},
  url = {https://link.aps.org/doi/10.1103/PhysRevLett.68.580}
}

@article{Brambilla:2024tqg,
    author = "Brambilla, Nora and Magorsch, Tom and Strickland, Michael and Vairo, Antonio and Vander Griend, Peter",
    title = "{Bottomonium suppression from the three-loop QCD potential}",
    eprint = "2403.15545",
    archivePrefix = "arXiv",
    primaryClass = "hep-ph",
    reportNumber = "TUM-EFT 186/23, FERMILAB-PUB-24-0029-V",
    doi = "10.1103/PhysRevD.109.114016",
    journal = "Phys. Rev. D",
    volume = "109",
    number = "11",
    pages = "114016",
    year = "2024"
}

@article{Strickland:2024oat,
    author = "Strickland, Michael and Thapa, Sabin and Vogt, Ramona",
    title = "{Bottomonium suppression in 5.02 and 8.16~TeV p-Pb collisions}",
    eprint = "2401.16704",
    archivePrefix = "arXiv",
    primaryClass = "nucl-th",
    doi = "10.1103/PhysRevD.109.096016",
    journal = "Phys. Rev. D",
    volume = "109",
    number = "9",
    pages = "096016",
    year = "2024"
}

@article{ParticleDataGroup:2026mpi,
    author = "Takahashi, F. and others",
    collaboration = "Particle Data Group",
    title = "{Review of Particle Physics$^\ast$}",
    doi = "10.1142/s0217751x26300115",
    journal = "Int. J. Mod. Phys. A",
    volume = "41",
    number = "22",
    pages = "2630011",
    year = "2026"
}

@article{CMS:2026qbb,
    author = "Belyaev, Andrey and others",
    collaboration = "CMS",
    title = "{Evidence for sequential $Υ$(nS) suppression in light ion collisions}",
    eprint = "2607.13758",
    archivePrefix = "arXiv",
    primaryClass = "nucl-ex",
    reportNumber = "CMS-HIN-25-015, CERN-EP-2026-188",
    month = "7",
    year = "2026"
}

@article{LHCb:2026jgc,
    author = "Aaij, Roel and others",
    collaboration = "LHCb",
    title = "{Modification of $Υ$ production in $p$O and OO collisions at LHCb}",
    eprint = "2608.00182",
    archivePrefix = "arXiv",
    primaryClass = "nucl-ex",
    reportNumber = "LHCb-PAPER-2026-027, CERN-EP-2026-207",
    month = "7",
    year = "2026"
}

@article{Matsui:1986dk,
    author = "Matsui, T. and Satz, H.",
    title = "{$J/\psi$ Suppression by Quark-Gluon Plasma Formation}",
    reportNumber = "BNL-38344",
    doi = "10.1016/0370-2693(86)91404-8",
    journal = "Phys. Lett. B",
    volume = "178",
    pages = "416--422",
    year = "1986"
}

@article{Andronic:2024oxz,
    author = "Andronic, A. and others",
    title = "{Comparative study of quarkonium transport in hot QCD matter}",
    eprint = "2402.04366",
    archivePrefix = "arXiv",
    primaryClass = "nucl-th",
    reportNumber = "FERMILAB-PUB-24-0005-T-V",
    doi = "10.1140/epja/s10050-024-01306-6",
    journal = "Eur. Phys. J. A",
    volume = "60",
    number = "4",
    pages = "88",
    year = "2024"
}

@article{Rothkopf:2019ipj,
    author = "Rothkopf, Alexander",
    title = "{Heavy Quarkonium in Extreme Conditions}",
    eprint = "1912.02253",
    archivePrefix = "arXiv",
    primaryClass = "hep-ph",
    doi = "10.1016/j.physrep.2020.02.006",
    journal = "Phys. Rept.",
    volume = "858",
    pages = "1--117",
    year = "2020"
}

@article{Yao:2021lus,
    author = "Yao, Xiaojun",
    title = "{Open quantum systems for quarkonia}",
    eprint = "2102.01736",
    archivePrefix = "arXiv",
    primaryClass = "hep-ph",
    reportNumber = "MIT-CTP/5273",
    doi = "10.1142/S0217751X21300106",
    journal = "Int. J. Mod. Phys. A",
    volume = "36",
    number = "20",
    pages = "2130010",
    year = "2021"
}

@article{Akamatsu:2020ypb,
    author = "Akamatsu, Yukinao",
    title = "{Quarkonium in quark{\textendash}gluon plasma: Open quantum system approaches re-examined}",
    eprint = "2009.10559",
    archivePrefix = "arXiv",
    primaryClass = "nucl-th",
    doi = "10.1016/j.ppnp.2021.103932",
    journal = "Prog. Part. Nucl. Phys.",
    volume = "123",
    pages = "103932",
    year = "2022"
}

@article{CMS:2023lfu,
    author = "Tumasyan, Armen and others",
    collaboration = "CMS",
    title = "{Observation of the {\Upsilon}(3S) Meson and Suppression of {\Upsilon} States in Pb-Pb Collisions at sNN=5.02{\,}{\,}TeV}",
    eprint = "2303.17026",
    archivePrefix = "arXiv",
    primaryClass = "hep-ex",
    reportNumber = "CMS-HIN-21-007, CERN-EP-2023-011",
    doi = "10.1103/PhysRevLett.133.022302",
    journal = "Phys. Rev. Lett.",
    volume = "133",
    number = "2",
    pages = "022302",
    year = "2024"
}

@article{CMS:2018zza,
    author = "Sirunyan, Albert M and others",
    collaboration = "CMS",
    title = "{Measurement of nuclear modification factors of $\Upsilon$(1S), $\Upsilon$(2S), and $\Upsilon$(3S) mesons in PbPb collisions at $\sqrt{s_{_\mathrm{NN}}} =$ 5.02 TeV}",
    eprint = "1805.09215",
    archivePrefix = "arXiv",
    primaryClass = "hep-ex",
    reportNumber = "CMS-HIN-16-023, CERN-EP-2018-110",
    doi = "10.1016/j.physletb.2019.01.006",
    journal = "Phys. Lett. B",
    volume = "790",
    pages = "270--293",
    year = "2019"
}

@article{Thapa:2026kok,
    author = "Thapa, Sabin and Wu, Biaogang and Vogt, Ramona and Rapp, Ralf",
    title = "{Comprehensive study of $Υ$($n$S) production in Oxygen+Oxygen collisions at LHC}",
    eprint = "2609.16292",
    archivePrefix = "arXiv",
    primaryClass = "nucl-th",
    month = "9",
    year = "2026"
}

\appendix
\clearpage
\onecolumngrid

\setcounter{equation}{0}
\setcounter{figure}{0}
\setcounter{table}{0}
\setcounter{page}{1}

\renewcommand{\theequation}{S\arabic{equation}}
\renewcommand{\thefigure}{S\arabic{figure}}
\renewcommand{\thetable}{S\arabic{table}}

\renewcommand{\theHfigure}{S\arabic{figure}}
\renewcommand{\theHtable}{S\arabic{table}}
\renewcommand{\theHequation}{S\arabic{equation}}

\renewcommand{\thepage}{S\arabic{page}}

\begin{center}
{\Large \textbf{Supplemental Material for ``System-size
dependence of bottomonium suppression from Pb-Pb to light-ion collisions''}}\\
Nora Brambilla, Tom Magorsch, Antonio Vairo
\end{center}

\twocolumngrid

\section{pNRQCD master equation}
\label{app:A}
We here provide the full form of the pNRQCD Lindblad equation~\eqref{eq:lind} in the spherical basis. As the density matrix is block diagonal in angular momentum $\rho_c=\bigoplus_\ell \rho^{\ell}_c$, we provide the Hamiltonian $H=\bigoplus_\ell H^{\ell}$ and jump operators $L_n$ in terms of angular momentum. The Hamiltonian $H^\ell$ is given by
\begin{align}
    H^\ell = \begin{pmatrix}
h^\ell_s + K_T & 0 \\
0 & h^\ell_o + \frac{N_{c}^{2}-2}{2\bigl(N_{c}^{2}-1\bigr)} K_T
\end{pmatrix},
\end{align}
with 
\begin{equation}
    K_T = \frac{r^{2}}{2}\gamma
+ \frac{\kappa}{4 M T}\{ r, p_{r} \}.
\end{equation}
Here $T$ is the temperature, $r$ is the radial-position operator, $p_r$ is its conjugate momentum operator, and $N_c=3$ in QCD. Following Ref.~\cite{Brambilla:2022ynh}, we fix the bottom-quark mass to $M=4.73\,\mathrm{GeV}$. The vacuum Hamiltonians $h_c^\ell$ are given by
\begin{equation}
    h^\ell_c = -\frac{1}{M}\left(\frac{\partial^2}{\partial r^2} + \frac{2}{r}\frac{\partial}{\partial r} \right) +V_c(r) + \frac{\ell(\ell+1)}{Mr^2},
\end{equation}
where $V_c(r)$ are the color-singlet and color-octet quarkonium potentials. At leading order these are given by attractive and repulsive Coulomb potentials
\begin{equation}
    V_s(r) = - \frac{C_F  \alpha_s}{r} , \qquad V_o (r) = \frac{\alpha_s}{2 N_c r}, 
\end{equation}
where $\alpha_s$ is the strong coupling. We use $\alpha_s=0.468$ based on the Bohr-radius prescription given in  Ref.~\cite{Brambilla:2022ynh}. 

The jump operators $L_n$ induce transitions between quantum numbers. The six different jump operators correspond to the transition singlet-to-octet, octet-to-singlet and octet-to-octet, each with the two possible angular momentum transitions $\ell\to\ell\pm 1$.
The jump operators are given by
\begin{align}
    &L_{0} = \sqrt{\frac{\kappa}{N^2_c-1}}\begin{pmatrix}
        0 & 1 \\ 0 & 0 
    \end{pmatrix}O^{\downarrow}C^\downarrow_{o\to s},\\
    &L_{1} = \sqrt{\frac{\kappa}{N^2_c-1}}\begin{pmatrix}
        0 & 1 \\ 0 & 0 
    \end{pmatrix}O^{\uparrow}C^\uparrow_{o\to s},\\
    &L_{2} = \sqrt{\kappa}\begin{pmatrix}
        0 & 0 \\ 1 & 0 
    \end{pmatrix}O^{\downarrow}C^\downarrow_{s\to o},\\
    &L_{3} = \sqrt{\kappa}\begin{pmatrix}
        0 & 0 \\ 1 & 0 
    \end{pmatrix}O^{\uparrow}C^\uparrow_{s\to o},\\
    &L_{4} = \sqrt{\frac{\kappa(N^2_c-4)}{2(N^2_c-1)}}\begin{pmatrix}
        0 & 0 \\ 0 & 1 
    \end{pmatrix}O^{\downarrow} C^\downarrow_{o\to o},\\
    &L_{5} = \sqrt{\frac{\kappa(N^2_c-4)}{2(N^2_c-1)}}\begin{pmatrix}
        0 & 0 \\ 0 & 1 
    \end{pmatrix}O^{\uparrow}C^\uparrow_{o\to o},
\end{align}
with operators in position space
\begin{align}
    &C^\downarrow_{u\to v}=r\left(1+\frac{\Delta V_{uv}}{4T}\right)+\frac{1}{2MT}\left(\frac{\partial}{\partial r}+\frac{\ell+1}{r}\right),\\
    &C^\uparrow_{u\to v}=r\left(1+\frac{\Delta V_{uv}}{4T}\right)+\frac{1}{2MT}\left(\frac{\partial}{\partial r}-\frac{\ell}{r}\right),
\end{align}
and $\Delta V_{uv}= V_u(r) - V_v(r)$. The operators $O^{\uparrow\downarrow}$ induce transitions between angular momentum blocks and are given by $O^\downarrow_{\ell^\prime,\ell} = \sqrt{\ell/(2\ell+1)} \delta_{\ell^\prime,\ell-1}$ and $O^\uparrow_{\ell^\prime,\ell} = \sqrt{(\ell+1)/(2\ell+1)} \delta_{\ell^\prime,\ell+1}$. 

In practice, we use the QTraj setup of Refs.~\cite{Brambilla:2023hkw,Brambilla:2022ynh}, which discretizes the radial wave function using $N=2048$ points in a box of size $L=40\,\mathrm{GeV}^{-1}$. We use the same compact Gaussian initial condition and evolve the wave function with a time step of $\delta t=\num{0.001}\,\mathrm{GeV}^{-1}$, as in previous studies. We impose an angular-momentum cutoff $\ell_{\mathrm{max}}$ and simulate only sectors with $\ell<\ell_{\mathrm{max}}$. If a trajectory reaches $\ell_{\mathrm{max}}$, we terminate it and set its contribution to the $S$- and $P$-wave overlaps to zero, since the probability of returning to a low-angular-momentum state is negligible. In our simulations, we use $\ell_{\mathrm{max}}=10$.

\section{Hydrodynamics}
\label{app:B}

The hydrodynamic evolution is calculated with aHydroQP~\cite{Alqahtani:2015qja,Alqahtani:2016rth,Alqahtani:2017mhy,Alqahtani:2020paa}, closely following the setup of previous bottomonium suppression studies performed for Pb-Pb collisions~\cite{Brambilla:2021wkt,Brambilla:2022ynh,Brambilla:2023hkw}. We take over this setup unchanged, including the Woods-Saxon parametrization of the nuclear density. Only the parameters that depend on the colliding nuclei and the beam energy are adjusted. In addition, we initialize the system with a non-vanishing momentum-space anisotropy.
\begin{figure}[!t]
      \centering
      \includegraphics[width=0.9\linewidth]{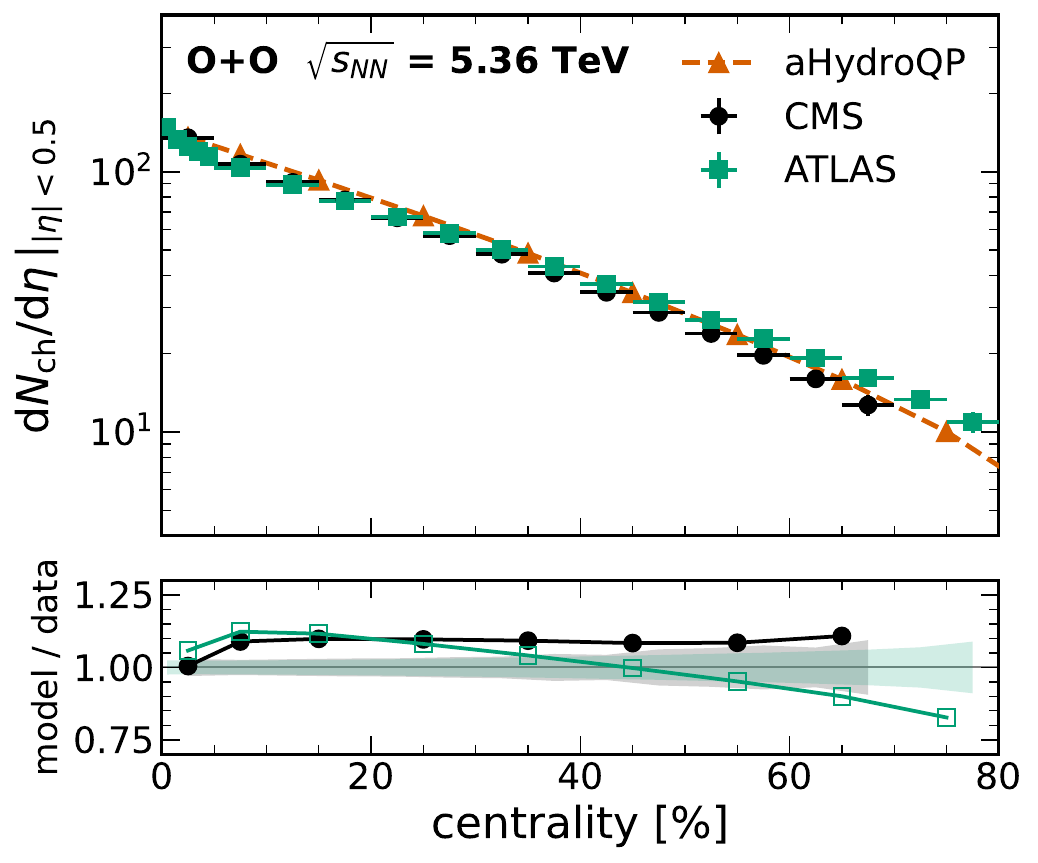}
      \caption{The midrapidity charged-particle density as a function of centrality, compared with data
  from the CMS~\cite{CMS:2026sai} and ATLAS~\cite{ATLAS:2026zgq} experiments.}
      \label{fig:hydrocent}
\end{figure}
\begin{figure}[!t]
    \centering
    \includegraphics[width=0.9\linewidth]{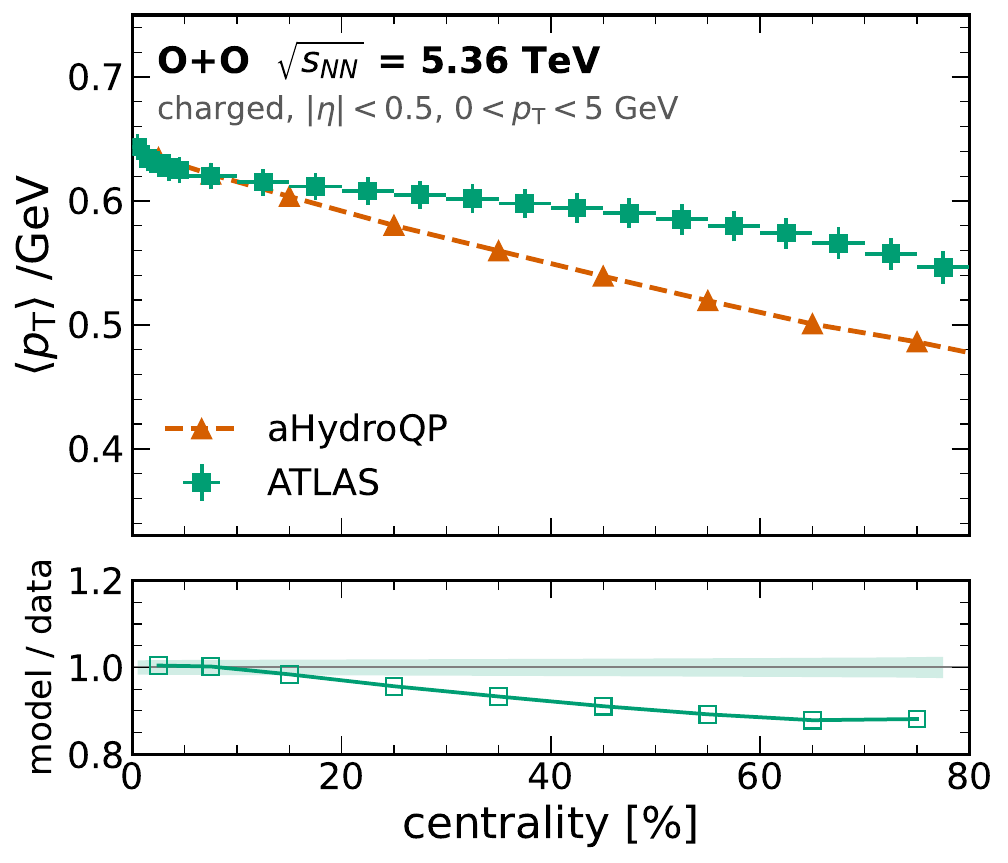}
    \caption{The transverse momentum $\braket{p_T}$ in terms of centrality. We compare the results from the aHydroQP simulations with data from the ATLAS experiment.}
    \label{fig:hydropT}
\end{figure}

For the surface diffuseness we use $a=0.513\,\mathrm{fm}$, from the three-parameter Fermi parametrization of Ref.~\cite{DEVRIES1987495}. In the aHydroQP setup the radius is fixed by the mass number, $R=1.12\,A^{1/3}-0.86\,A^{-1/3}$, which we retain. For $^{16}$O this gives $R=2.481\,\mathrm{fm}$. The saturation density $\rho_0=0.1759\,\mathrm{fm}^{-3}$ is fixed by $\int\rho\,{\rm d}^3r=A$. 
The inelastic nucleon-nucleon cross section is set to $\sigma_{NN}^\mathrm{inel}=68\,\mathrm{mb}$ at $\sqrt{s_{NN}}=5.36\,\mathrm{TeV}$~\cite{Loizides:2025ule}. The parameters of the initial condition, the binary-collision admixture $\chi=0.12$ and the plateau and Gaussian half-widths $\Delta_\varsigma=3.0$ and $\sigma_\varsigma=2.0$ of the longitudinal profile  are fitted together with $T_0$ to the measured pseudorapidity distribution shown in the main text. 
Centrality classes are obtained from the same optical-Glauber calculation, with each class represented by a single impact parameter.
The evolution uses the HotQCD equation of state~\cite{HotQCD:2014kol} in the quasiparticle implementation of Ref.~\cite{Alqahtani:2015qja}.
The freeze-out hypersurface is constructed at a decoupling temperature of $T_{\rm FO}=150\,\mathrm{MeV}$. The values  $\alpha_x=\alpha_y=1$, $\alpha_z=0.4$ are motivated by the free-streaming limit of the boost-invariant longitudinal expansion. Starting from an approximately isotropic distribution at a microscopic time $\tau_{\rm micro}$, free streaming generates an anisotropy $\xi_{\rm FS}(\tau)=(\tau/\tau_{\rm micro})^2-1$~\cite{Martinez_2008}.
For the spheroidal parametrization used here, $\alpha_z=(1+\xi)^{-1/2}$, such that $\alpha_z(\tau)=\tau_{\rm micro}/\tau$.
Taking $\tau_{\rm micro}\simeq 0.1\,{\rm fm}$~\cite{Strickland_2014}, gives $\alpha_z(\tau_0=0.25\,{\rm fm})\simeq0.4$.
We use this estimate only to set the characteristic size of the initial deformation and do not attempt to model the pre-equilibrium evolution explicitly. With $\alpha_z\neq1$ the initial distribution is not thermal, and the effective temperature obtained by Landau matching no longer coincides with the momentum scale $\lambda$ of the distribution. For $T_0=466\,\mathrm{MeV}$ and $\alpha_z=0.4$ we find $\lambda_0=601\,\mathrm{MeV}$.
Beyond the centrality-integrated distribution shown in the main text, we compare the model to data differentially in centrality, for the charged-particle multiplicity and the mean transverse momentum.
\begin{figure}[!t]
      \centering
      \includegraphics[width=0.9\linewidth]{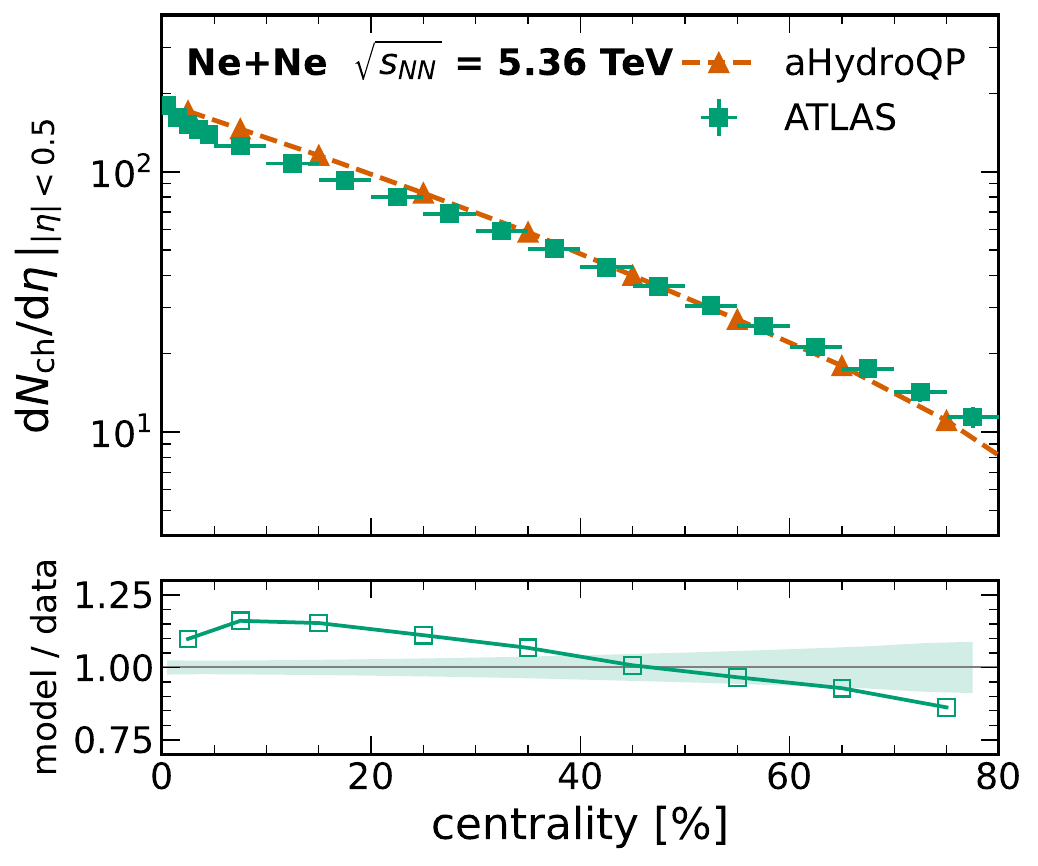}
      \caption{The midrapidity charged-particle multiplicity as a function of centrality, compared with data
  from the ATLAS~\cite{ATLAS:2026zgq} experiment for Ne-Ne collisions.}
      \label{fig:hydrocentne}
\end{figure}
\begin{figure*}[!t]
    \centering
    \includegraphics[width=0.9\linewidth]{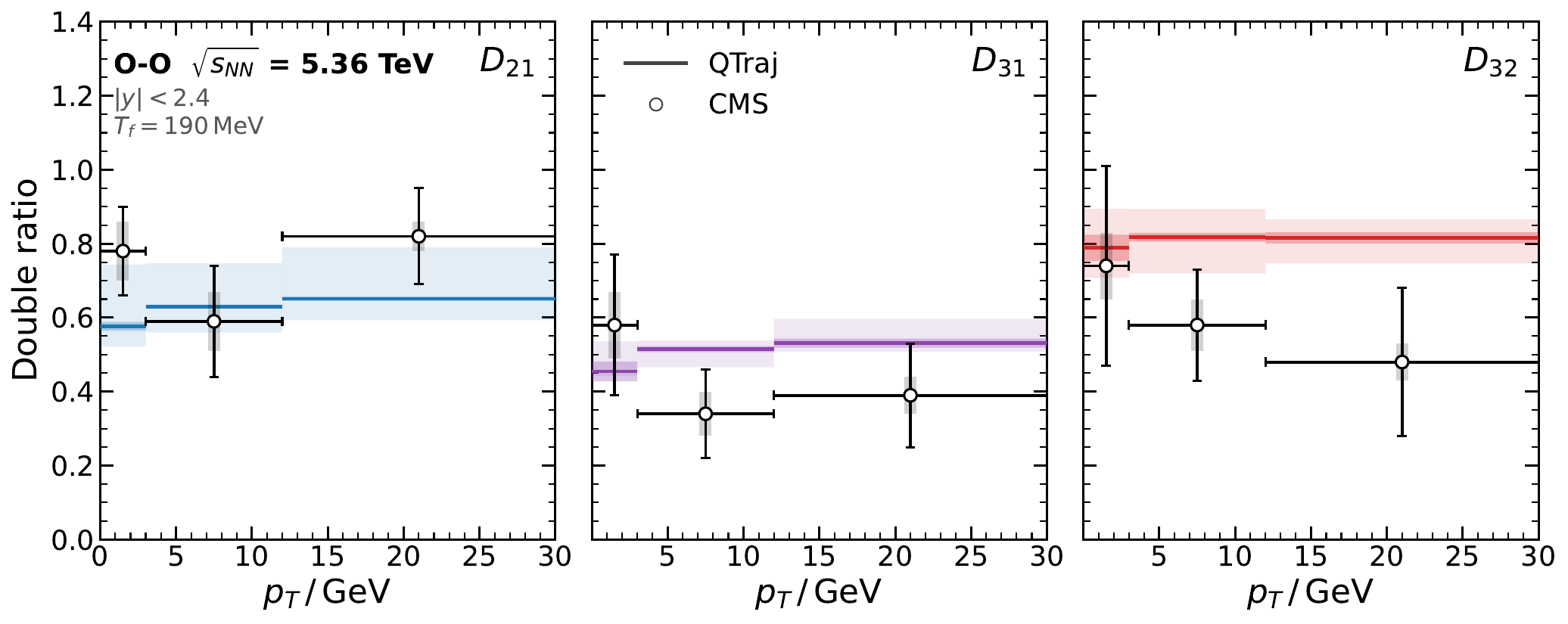}
    \caption{The double ratios $D_{ij}$ in terms of transverse momentum $p_T$ for O-O collisions compared to CMS measurements~\cite{CMS:2026qbb}. The central QTraj prediction corresponds to the transport coefficients $\hat\kappa=4$ and $\hat\gamma=0$ and the final temperature $T_f=190\,$MeV. The light shaded bands correspond to the independent variation of $\hat\kappa$ and $\hat\gamma$ by $\pm 1$. The dark shaded band around the central QTraj prediction corresponds to the statistical uncertainty from the Monte Carlo sampling.}
    \label{fig:pT}
\end{figure*}

In Fig.~\ref{fig:hydrocent} we compare the midrapidity charged-particle density as a function of centrality with the CMS and ATLAS measurements. CMS and ATLAS differ from each other toward peripheral centrality, and neither experiment reports results for the most peripheral centrality classes. Centrality determination is known to become increasingly uncertain toward peripheral collisions~\cite{CMS:2026sai}. 
The aHydroQP calculation reproduces CMS in the most central class but overshoots the CMS result for higher centralities.
Similarly, it is larger than the ATLAS measurement for central collisions, while it becomes significantly smaller for peripheral collisions, thus effectively sitting between the CMS and ATLAS measurements above $50\%$ centrality. Several factors likely contribute to this residual discrepancy, including the difference between classifying events by impact parameter and by an experimental centrality estimator, and the more general limitations of a smooth, fluctuation-free initial condition.

In Fig.~\ref{fig:hydropT} we compare the calculated mean transverse momentum as a function of centrality with the ATLAS measurement. The calculation reproduces the data in the most central classes, but increasingly underestimates $\langle p_T\rangle$ toward peripheral collisions. Previous studies of Pb-Pb collisions, where relative fluctuations are much smaller than in O-O, have found inclusive bottomonium suppression to be only weakly modified by event-by-event fluctuations of the hydrodynamic background~\cite{Alalawi:2022gul}. Whether this remains true for a system as small as O-O is not established.

For the hydrodynamic evolution of Ne-Ne collisions, we use the same setup with an initial central effective temperature of $T_0^{\mathrm{Ne}}=481\,\mathrm{MeV}$. We show the charged-particle multiplicity as a function of centrality in Fig.~\ref{fig:hydrocentne}.

\section{Transverse-momentum-dependent double ratios}

In Fig.~\ref{fig:pT} we show the $p_T$ distribution of the double ratios in O-O collisions computed from QTraj for the final temperature $T_f=190\,$MeV compared to the CMS measurements. For $D_{21}$ and $D_{31}$ the double ratios show a slow increase with $p_T$, which may arise from a shorter time spent by faster bottomonia in the QGP. This trend is not visible in the CMS data, although the large uncertainties prevent stronger conclusions. In $D_{32}$ the slight increase of the two $p_T$ distributions largely cancels, leading to an approximately flat prediction.

\end{document}